# Improvement of Mixing Function for Modified Upwinding Compact Scheme

Huankun Fu[1], Ping Lu[2],
UNIVERSITY OF TEXAS AT ARLINGTON, ARLINGTON, TX 76019, USA

Chaoqun Liu[3]
UNIVERSITY OF TEXAS AT ARLINGTON, ARLINGTON, TX 76019, USA
CLIU@UTA.EDU

The compact scheme has high order accuracy and high resolution, but cannot be used to capture the shock. WENO is a great scheme for shock capturing, but is too dissipative for turbulence and small length scales. We developed a modified upwinding compact scheme which uses an effective shock detector to block compact scheme to cross the shock and a control function to mix the flux with WENO scheme near the shock. The new scheme makes the original compact scheme able to capture the shock sharply and, more important, keep high order accuracy and high resolution in the smooth area, which is particularly important for shock boundary layer and shock acoustic interactions. This work is a continuation to modify the control function for the modified up-winding compact scheme (MUCS). Numerical results show the scheme is successful for 2-D Euler.

## I. Introduction

THIS Compact scheme (Lele, 1992) is a great scheme which has high order accuracy and high resolution, which is effective in DNS/LES for the turbulence simulation. However, compact scheme cannot be used for shock capturing since the calculation of derivatives requires use of both downstream and upstream points. The WENO (Jiang & Shu, 1996) scheme is a great scheme for shock capturing with a third order accuracy near the shock and $5^{th}$ order accuracy away of shock. However, WENO is still too dissipative for DNS/LES for the turbulence simulation. A combination of the compact scheme and WENO scheme should be desirable. There are some efforts to combine WENO with upwinding compact (UCS) scheme (Ren et al, 2003). However, their mixing function is still some kind complex and has a number of case related adjustable coefficients.

Last year, we use WENO to improve $7^{th}$ order upwinding compact scheme as we called as "modified upwinding compact scheme (MUCS)", which uses a new shock detector to find the shock location and a new control function to mix upwinding compact scheme with WENO. The mixing function is designed in following ways: the new scheme automatically becomes bias when approaching the shock, but rapidly recovers to be upwinding compact, with high order of accuracy and high resolution.

However, the mixing function must be optimized for high efficiency. It is required that the mixing function must be smooth (not a switch function), keeps up-winding for shock, keeps enough dissipation before and after shock, and maintain high accuracy in the smooth region.

## II. Compact and WENO Schemes

### 2.1 Compact Scheme
Before discussing our new scheme, first let us see how to construct the CS and WENO schemes.

---

[1],[2] PhD Students, Math department, 411 Nedderman Drive, Pickard Hall, Arlington, TX
[3] Professor, Math department, 411 Nedderman Drive, Pickard Hall, Arlington, TX

### 2.1.1 Primitive function for conservation

For 1-D conservation laws:

$$u_t(x,t) + f_x(u(x,t)) = 0 \tag{2.1}$$

When a conservative approximation to the spatial derivative is applied, a semi-discrete conservative form of the equation (2.1) is described as follows:

$$\frac{du_j}{dt} = -\frac{1}{\Delta x}(\hat{f}_{j+(1/2)} - \hat{f}_{j-(1/2)}) \tag{2.2}$$

where $f_j = \frac{1}{\Delta x}\int_{x_j-\Delta x/2}^{x_j+\Delta x/2} \hat{f}(\xi)d\xi$, then $(f_x)_j = -\frac{1}{\Delta x}(\hat{f}_{j+(1/2)} - \hat{f}_{j-(1/2)})$. Note that f is the original function, but $\hat{f}$ is the flux defined by the above integration which is an exact formula of the flux but is different from f.

Let H be the primitive function of $\hat{f}$ defined below:

$$H(x_{j+(1/2)}) = \int_{-\infty}^{x_j+\Delta x/2} \hat{f}(\xi)d\xi = \sum_{i=-\infty}^{i=j} \int_{x_i-\Delta x/2}^{x_i+\Delta x/2} \hat{f}(\xi)d\xi = \Delta x \sum_{i=-\infty}^{j} f_i \tag{2.3}$$

H is easy to be calculated, but is a discrete data set.

The numerical flux $\hat{f}$ at the cell interfaces is the derivative of its primitive function H. i.e.:

$$\hat{f}_{j+(1/2)} = H'_{j+(1/2)} \tag{2.4}$$

All formulae given above are exact without approximations. However, the primitive function H is a discrete data set or discrete function and we have to use numerical method to get the derivatives, which will introduce numerical errors, or, in other words, order of accuracy.

This procedure, $f \rightarrow H \rightarrow \hat{f} \rightarrow f'_x$, was introduced by Shu & Osher (1988, 1989). There is still one problem left for numerical methods, which is how to solve (2.4) or how to get accurate derivatives for a discrete data set.

### 2.1.2 High-order compact schemes

A Pade-type compact scheme with a one parameter can be written (Lele, 1992):

$$\alpha f'_{i-1} + f'_i + \alpha f'_{i+1} = \left[-\frac{1}{12}(4\alpha-1)f_{i-2} - \frac{1}{3}(\alpha+2)f_{j-1} + \frac{1}{3}(\alpha+2)f_{j+1} + \frac{1}{12}(4\alpha-1)f_{i+2}\right]/h \tag{2.5}$$

If $\alpha = \frac{1}{3}$, we will get a standard sixth order compact scheme

$$\frac{1}{3}f'_{i-1} + f'_i + \frac{1}{3}f'_{i+1} = \left[-\frac{1}{36}f_{i-2} - \frac{7}{9}f_{j-1} + \frac{7}{9}f_{j+1} + \frac{1}{36}f_{i+2}\right]/h \tag{2.6}$$

Note that we only use CS for primitive function to get flux:

$$\hat{f}_{j+(1/2)} = H'_{j+(1/2)}$$

### 2.1.3 Upwinding Compact Scheme

The standard compact scheme does not have dissipation (non-dissipative scheme) and needs filter even for smooth area. The upwinding compact scheme can keep the high order without the filter. Following our conservative primitive function approach, the 7$^{th}$ order up-winding scheme can be described as follows:

For the positive primitive function $H^+$:

$$\frac{1}{2}H'_{j-\frac{3}{2}} + H'_{j-\frac{1}{2}} + \frac{1}{4}H'_{j+\frac{1}{2}} = \frac{\left(\frac{1}{240}H_{j-\frac{7}{2}} - \frac{1}{12}H_{j-\frac{5}{2}} - \frac{11}{12}H_{j-\frac{3}{2}} + \frac{1}{3}H_{j-\frac{1}{2}} + \frac{31}{48}H_{j+\frac{1}{2}} + \frac{1}{60}H_{j+\frac{3}{2}}\right)}{h} \tag{2.7}$$

For the negative primitive function $H^-$:

$$\frac{1}{4}H'_{j-\frac{3}{2}} + H'_{j-\frac{1}{2}} + \frac{1}{2}H'_{j+\frac{1}{2}} = \frac{\left(-\frac{1}{60}H_{j-\frac{5}{2}} - \frac{31}{48}H_{j-\frac{3}{2}} - \frac{1}{3}H_{j-\frac{1}{2}} + \frac{11}{12}H_{j+\frac{1}{2}} + \frac{1}{12}H_{j+\frac{3}{2}} - \frac{1}{240}H_{j+\frac{5}{2}}\right)}{h} \quad (2.8)$$

Here, $H^+$ represents primitive function of positive flux and $H^-$ represents primitive function of negative flux. A Lax-Friedrich flux splitting is applied in all computational examples.

## 2.2 WENO Scheme (Jiang & Su, 1996)

The basic idea proposed in ENO (Harten et al, 1987) and WENO (Jiang et al, 1996) schemes is to avoid the stencil containing a shock. ENO chooses the smoothest stencil from several candidates to calculate the derivatives. WENO controls the contributions of different stencils according to their smoothness. In this way, the derivative at a certain grid point, especially one near the shock, is dependent on a very limited number of grid points. The local dependency here is favorable for shock capturing and helps obtaining the non-oscillatory property. The success of ENO and WENO schemes indicates that the local dependency is critical for shock capturing.

### 2.2.1 Conservation Form of Derivative

$$\frac{\partial U}{\partial t} + \frac{\partial F}{\partial x} = 0 \quad (2.9)$$

The ENO reconstruction can provide a semi-discretization for the derivative: $\frac{\partial F}{\partial x} = \frac{\hat{F}_{i+\frac{1}{2}} - \hat{F}_{i-\frac{1}{2}}}{\Delta x}$, where $\hat{F}$ is the flux which must be accurately obtained.

### 2.2.2 5$^{th}$ Order WENO (bias upwind)

*1. Flux approximation*

In order to get an high order approximation for $\hat{F}_{j-\frac{1}{2}} = H'_{j-\frac{1}{2}}$, we can use three different candidates (Figure 2.1) which are all third order polynomials: $E_0 : H_{j-\frac{7}{2}}, H_{j-\frac{5}{2}}, H_{j-\frac{3}{2}}, H_{j-\frac{1}{2}}$; $E_1 : H_{j-\frac{5}{2}}, H_{j-\frac{3}{2}}, H_{j-\frac{1}{2}}, H_{j+\frac{1}{2}}$;

$E_2 : H_{j-\frac{3}{2}}, H_{j-\frac{1}{2}}, H_{j+\frac{1}{2}}, H_{j+\frac{3}{2}}$.

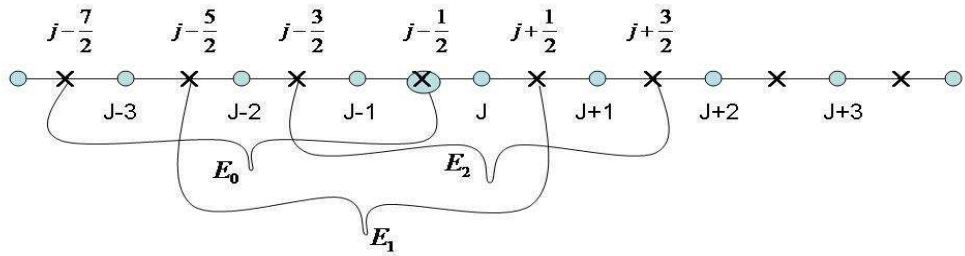

(a)

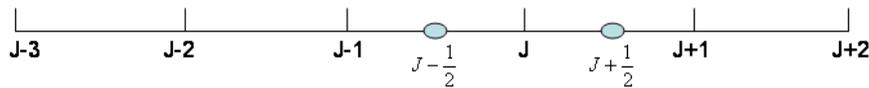

(b)

**Figure 2.1** (a) WENO candidates (b) 5$^{th}$ order WENO Scheme

Let us look at candidate $E_0$ first. Assume H is a third order polynomial:

3
American Institute of Aeronautics and Astronautics

$H = a_0 + a_1(x - x_{j-1/2}) + a_2(x - x_{j-1/2})^2 + a_3(x - x_{j-3/2})^3$, we have

$$H_{j-1/2} = a_0$$
$$H_{j-3/2} = a_0 - a_1 h + a_2 h^2 - a_3 h^3$$
$$H_{j-5/2} = a_0 - 2a_1 h + 4a_2 h^2 - 8a_3 h^3 \quad (2.10)$$
$$H_{j-7/2} = a_0 - 3a_1 h + 9a_2 h^2 - 27 a_3 h^3$$

Finally, we have

$$E_0 : \hat{F}_{j-\frac{1}{2}} = \frac{1}{3} F_{j-3} - \frac{7}{6} F_{j-2} + \frac{11}{6} F_{j-1}$$
$$E_1 : \hat{F}_{j-\frac{1}{2}} = -\frac{1}{6} F_{j-2} + \frac{5}{6} F_{j-1} + \frac{1}{3} F_j \quad (2.11)$$
$$E_2 : \hat{F}_{j-\frac{1}{2}} = \frac{1}{3} F_{j-1} + \frac{5}{6} F_j - \frac{1}{6} F_{j+1}$$

2. *Optimal weights for high order of accuracy*

The final scheme should be a combination of three candidates: $E = C_0 E_0 + C_1 E_1 + C_2 E_2$.

*If we set* $C_0 = \frac{1}{10}, C_1 = \frac{6}{10}, C_2 = \frac{3}{10}$, *we will have*

$$\hat{F}_{j-\frac{1}{2}} = \frac{1}{30} F_{j-3} - \frac{13}{60} F_{j-2} + \frac{47}{60} F_{j-1} + \frac{27}{60} F_j - \frac{1}{20} F_{j+1},$$
$$\hat{F}_{j+\frac{1}{2}} = \frac{1}{30} F_{j-2} - \frac{13}{60} F_{j-1} + \frac{47}{60} F_j + \frac{27}{60} F_{j+1} - \frac{1}{20} F_{j+2}, \quad (2.12)$$
$$\frac{\partial F}{\partial x} = \frac{\hat{F}_{j+\frac{1}{2}} - \hat{F}_{j-\frac{1}{2}}}{\Delta x} = (-\frac{1}{30} F_{j-3} + \frac{1}{4} F_{j-2} - F_{j-1} + \frac{1}{3} F_j + \frac{1}{2} F_{j+1} - \frac{1}{20} F_{j+2})/\Delta x + O(\Delta x^5)$$

Using Taylor expansion for $F_{j-k}$, we find

$$\frac{\partial F}{\partial x} = \frac{\hat{F}_{j+\frac{1}{2}} - \hat{F}_{j-\frac{1}{2}}}{\Delta x} = F'_j - \frac{1}{60}(\Delta x)^5 F_j^{(6)} + \frac{1}{140}(\Delta x)^6 F_j^{(7)} + \ldots, \quad (2.13)$$

which shows the scheme with optimal weights and 6 grid points has a $5^{th}$ order truncation error. Note that the scheme is a **STANDARD** $5^{th}$ order bias upwind finite difference scheme.

3. *Bias up-wind weights:*

Let us define a bias weight for each candidate according to WENO:

$$\omega_k = \frac{\gamma_k}{\sum_{i=0}^{2} \gamma_i}, \quad \gamma_k = \frac{C_k}{(\varepsilon + IS_k)^p} \quad (2.14)$$

where

$$IS_i = \int_{x_{j-1/2}}^{x_{j+1/2}} \sum_{k=1}^{\infty} [p_2(x)^{(k)}]^2 h^{2k-1} dx$$

$$IS_0 = \frac{13}{12}(F_{j-2} - 2F_{j-1} + F_j)^2 + \frac{1}{4}(F_{j-2} - 4F_{j-1} + 3F_j)^2$$

$$IS_1 = \frac{13}{12}(F_{j-1} - 2F_j + F_{j+1})^2 + \frac{1}{4}(F_{j-1} - F_{j+1})^2$$

$$IS_2 = \frac{13}{12}(F_j - 2F_{j+1} + F_{j+2})^2 + \frac{1}{4}(F_{j+2} - 4F_{j+1} + 3F_j)^2$$

The $5^{th}$ order WENO can be obtained

$$\hat{F}_{j-1/2} = \omega_0 E_0 + \omega_1 E_1 + \omega_2 E_2 \tag{2.15}$$

$$\hat{F}_{j-1/2} = \omega_{0,j-1/2}(\frac{1}{3}F_{j-3} - \frac{7}{6}F_{j-2} + \frac{11}{6}F_{j-1}) + \omega_{1,j-1/2}(-\frac{1}{6}F_{j-2} + \frac{5}{6}F_{j-1} + \frac{1}{3}F_j)$$
$$+ \omega_{2,j-1/2}(\frac{1}{3}F_{j-1} + \frac{5}{6}F_j - \frac{1}{6}F_{j+1}) \tag{2.16}$$

WENO is a great scheme with great successes by many users. However, the scheme has 5th order dissipation everywhere and third order dissipation near the shock and people in DNS/LES community complain it is too dissipative for transition and turbulence. Let us turn into compact schemes for assistance.

## III. Modified Compact Scheme

Compact scheme is great to resolve small length scales, but cannot be used for the cases when a shock or discontinuity is involved. Our new modified compact scheme is an effort to remove the weakness by introducing WENO flux when the computation is approaching the shock.

### 3.1 Basic Idea of the Control Function

Although the new shock detector can provide accurate location of shock including weak shock, strong shock, oblique shock and discontinuity in function, first, second and third order derivatives, it is a switch function and give one in shock and zero for others. As we mentioned above, a switch function cannot be directly used to mix CS and WENO and we must develop a rather smooth function to mix CS (3.9) and WENO (3.10):

$$\frac{1}{3}H'_{j-\frac{3}{2}} + H'_{j-\frac{1}{2}} + \frac{1}{3}H'_{j+\frac{1}{2}} = \left[-\frac{1}{36}H_{j-\frac{5}{2}} - \frac{7}{9}H_{j-\frac{3}{2}} + \frac{7}{9}H_{j+\frac{1}{2}} + \frac{1}{36}H_{j+\frac{3}{2}}\right]/h \tag{3.1}$$

$$H'_{j-1/2} = \omega_{0,j-1/2}(\frac{1}{3}F_{j-3} - \frac{7}{6}F_{j-2} + \frac{11}{6}F_{j-1}) + \omega_{1,j-1/2}(-\frac{1}{6}F_{j-2} + \frac{5}{6}F_{j-1} + \frac{1}{3}F_j)$$
$$+ \omega_{2,j-1/2}(\frac{1}{3}F_{j-1} + \frac{5}{6}F_j - \frac{1}{6}F_{j+1}) \tag{3.2}$$

where F is the original function and H is a primitive function of F and $H'$ is the flux $\hat{F} = H'$.

We defined a new control function $\Omega$:
$$(1-\Omega)*CS + \Omega*WENO \tag{3.3}$$

This will lead a tri-diagonal matrix system which is the core of our new scheme:

$$\frac{1}{3}(1-\Omega)H'_{j-\frac{3}{2}} + H'_{j-\frac{1}{2}} + \frac{1}{3}(1-\Omega)H'_{j+\frac{1}{2}} = (1-\Omega)[\frac{1}{36}(H_{j+\frac{3}{2}} - H_{j-\frac{5}{2}}) + \frac{7}{9}(H_{j+\frac{1}{2}} - H_{j-\frac{3}{2}})]/h$$
$$+\Omega*[\omega_{0,j-1/2}(\frac{1}{3}F_{j-3} - \frac{7}{6}F_{j-2} + \frac{11}{6}F_{j-1}) + \omega_{1,j-1/2}(-\frac{1}{6}F_{j-2} + \frac{5}{6}F_{j-1} + \frac{1}{3}F_j)$$
$$+\omega_{2,j-1/2}(\frac{1}{3}F_{j-1} + \frac{5}{6}F_j - \frac{1}{6}F_{j+1})] \tag{3.4}$$

When $\Omega = 0.0$, the equation become a standard sixth order compact scheme, but when $\Omega = 1.0$ the scheme is a standard WENO scheme.

For the modified upwinding compact scheme (MUCS), the final matrix becomes:

$$\frac{1}{4}(1-\Omega)H'_{j-\frac{3}{2}} + H'_{j-\frac{1}{2}} + \frac{1}{2}(1-\Omega)H'_{j+\frac{1}{2}} = (1-\Omega)*[-\frac{1}{60}H_{j-\frac{5}{2}} - \frac{31}{48}H_{j-\frac{3}{2}} - \frac{1}{3}H_{j-\frac{1}{2}} + \frac{11}{12}H_{j+\frac{1}{2}} + \frac{1}{12}H_{j+\frac{3}{2}}$$
$$-\frac{1}{240}H_{j+\frac{5}{2}}]/h + \Omega*[\omega_{0,j-1/2}(\frac{1}{3}F_{j-3} - \frac{7}{6}F_{j-2} + \frac{11}{6}F_{j-1}) + \omega_{1,j-1/2}(-\frac{1}{6}F_{j-2} + \frac{5}{6}F_{j-1} + \frac{1}{3}F_j)$$
$$+\omega_{2,j-1/2}(\frac{1}{3}F_{j-1} + \frac{5}{6}F_j - \frac{1}{6}F_{j+1})] \tag{3.5}$$

### 3.2 Construction of the Control Function

In our new shock detector, we define $MR(i,h) = \dfrac{T_C(i,h)}{T_F(i,h) + \varepsilon}$ as a ratio of coarse grid truncation over find grid truncation. That is,

$$T_C(i,h) = T_4(i,2h) + T_5(i,2h) + T_6(i,2h) = \frac{|f_i^{(4)}|(2h)^4}{4!} + \frac{|f_i^{(5)}|(2h)^5}{5!} + \frac{|f_i^{(6)}|(2h)^6}{6!} \text{ and}$$

$$T_F(i,h) = T_4(i,h) + T_5(i,h) + T_6(i,h) = \frac{|f_i^{(4)}|(h)^4}{4!} + \frac{|f_i^{(5)}|(h)^5}{5!} + \frac{|f_i^{(6)}|(h)^6}{6!} \tag{3.6}$$

MR should be around 16.0 if the function has at least $6^{th}$ order continuous derivatives.

The new local left and right slope ratio check is:

$$LR(i) = \frac{\left\| \frac{f'_R}{f'_L} \right| - \left| \frac{f'_L}{f'_R} \right\|}{\left| \frac{f'_R}{f'_L} \right| + \left| \frac{f'_L}{f'_R} \right| + \varepsilon} = \left| \frac{(f'_R)^2 - (f'_L)^2}{(f'_R)^2 + (f'_L)^2 + \varepsilon} \right| = \left| \frac{\alpha_R^2 - \alpha_L^2}{\alpha_R^2 + \alpha_L^2 + \varepsilon} \right| \tag{3.7}$$

Where $f'_R = 3f_i - 4f_{i+1} + f_{i+2}$, $f'_L = 3f_i - 4f_{i-1} + f_{i-2}$ and $\varepsilon$ is a small number to avoid division by zero.

### 3.2.1 Original Control Function

Our origin control function is

$$\Omega = \min[1.0, \ 8.0/MR(i,h)] \times LR(i,h) \tag{3.8}$$

in which $MR$ is the multigrid global truncation error ratio and $LR$ is local ratio of left and right side angle ratio. If the shock is met, MR is small, LR is near 1 and $\Omega = 1.0$, the WENO will be used and the CS is fully blocked. If the area is smooth, MR should be around 16.0 and LR is close to zero (left and right angle are same). Additional requirement is set that any point must compare with left and right neighboring points and we pick the largest $\Omega$ among the three neighboring points.

The reason we pick 8.0 is that we treat the fourth order continuous function as smooth function and only need half of LR for $\Omega$. It is easy to find there are no case related adjustable coefficients which is quite different from many other published hybrid schemes. However, as the mixing function, sometimes its value is too small for the shock location, and the scheme smears too much. Our new control function is better.

### 3.2.2 New Control function:

We define

$$A(i,h) = \sqrt{\min[1.0, \ 4.0/MR(i,h)] \times LR(i,h)}, \tag{3.9}$$

We set $\Omega = (A(i-1,h) + A(i,h) + A(i+1,h))/3.0$

For $A(i,h)$, we have the square root because $\min[1.0, 4.0/MR(i,h)] \times LR(i,h)$ is the product of two values, both of which are smaller than one. The consequent value becomes too small for the shock area, and therefore we use the square root to "recover" the value to be near 1.0 as much as we can. We use the average of the three consecutive values as the final weight of WENO because the average can reduce the possibility of misjudgments and makes the control function much smoother. The following is the result of using our new control function, and we made some comparison between the new scheme and the pure WENO.

## IV. Computational Results by New MUCS

**New MUCS for 2-D Euler Equations**

An incident shock case with an inflow Mach number of 2 and attach angle of $\vartheta = 35.241^0$ was chosen as a sample problem to compare the WENO and MUCS results with the exact solution. Since the incident shock has exact solution, it is a good prototype problem for scheme validation and comparison. It is also a difficult problem to get sharp shock without visible oscillation for any high order scheme since it has oblique shocks involved. The computational domain is x=2.0 and y=1.1 and a uniform grids was used. We find that modified compact scheme (MCS) worked well on coarse and middle size grids, but has oscillations on the fine grids (129x129). While modified upwinding compact scheme or MUCS does not have serious oscillation, even better than WENO after the second shock. On the other hand MUCS captured shock sharper than WENO for all grids. The control (mixing) function did use WENO (red and yellow: WENO dominated) to block the UCS and used UCS for smooth area (blue area: UCS dominated). All of the comparisons are made by using same code and same boundary treatment but different subroutines (WENO or MUCS) for derivatives only. The fine grids (129×129) results are depicted on Figures 4.1-43. From these figures we can find the 7$^{th}$ order MUCS results are very comparable with exact solution and are better than that obtained by 5$^{th}$ order WENO scheme. Figure 4.2 (b) shows that our shock detector works pretty well and captures the shock accurately. Our new control function (mixing function) also gives good weights for WENO and UCS, which indicates that when the shock is met, WENO becomes dominant gradually, but in smooth area, the scheme is dominated by UCS. Figures 4.3 (a) and (b) give us the comparison of pressure on the wall between our numerical solution and exact solution, which shows that our result is very near the exact solution although it is a little overshooting after the second shock. In Figures 4.3 (c), (d), (e), (f) and (g), the comparisons of pressure on the wall and K=30 between our new scheme and pure WENO are given. Figures 4.3 (e), (f), (g) are locally enlarged for comparison. As seen, for pressure on the wall, although both of our new scheme and WENO have a little overshooting and oscillation immediately after the second shock (Figure 4.3 (f)), the pure WENO smears flow too much. This smearing also occurs in all other level, e.g. K=30 (Figures 4.3 (d) and (g)). We know the smearing should be avoided for small length scales, especially for turbulence. Therefore, our new scheme is much more compatible for small length scales.

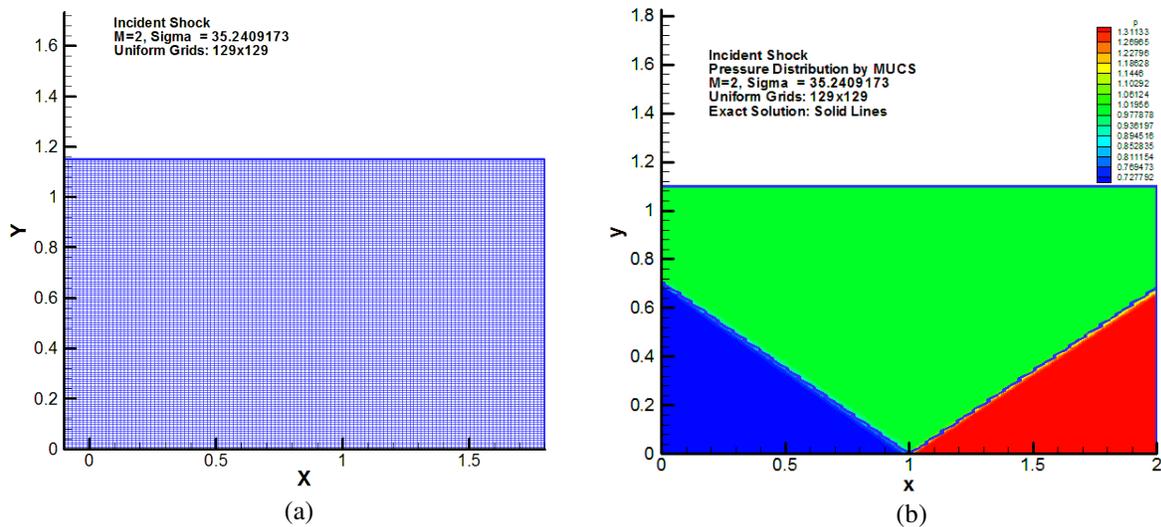

**Figure 4.1 Numerical test for 2D incident shock on fine grids** (a) Grids (129x129) (b) Pressure contour

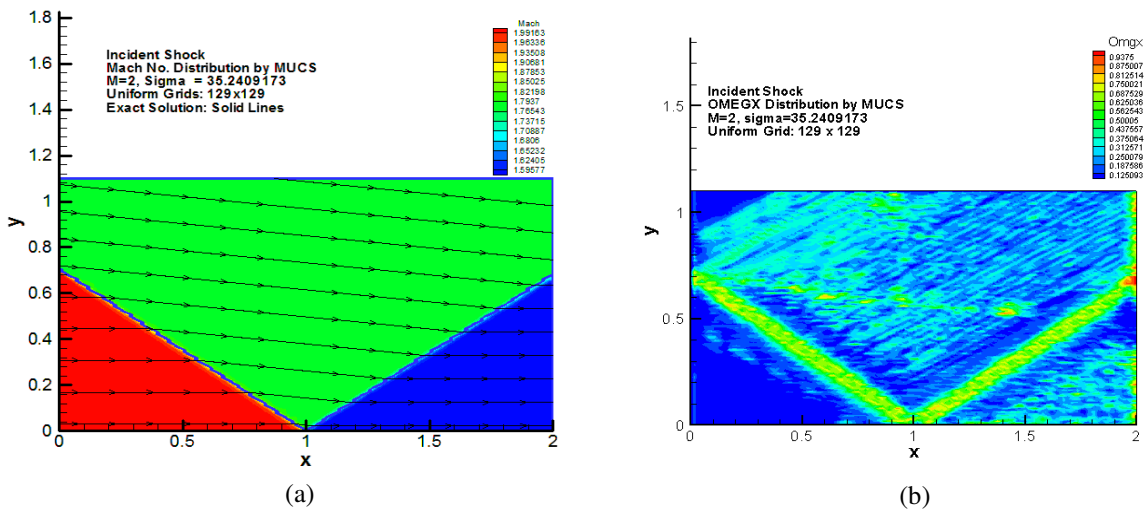

**Figure 4.2 Numerical test for 2-D incident shock on fine grids** (a) Mach number
(b) Control function (red and yellow: WENO dominated; blue: UCS dominated)

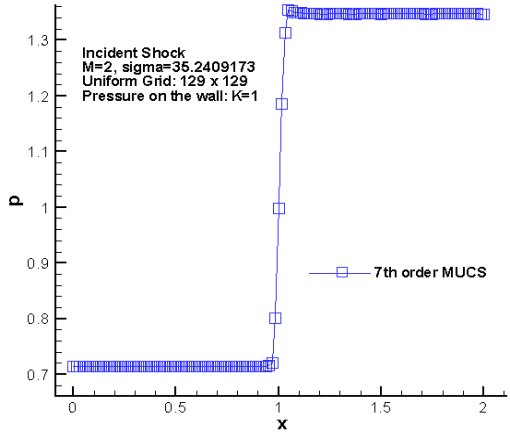

(a)

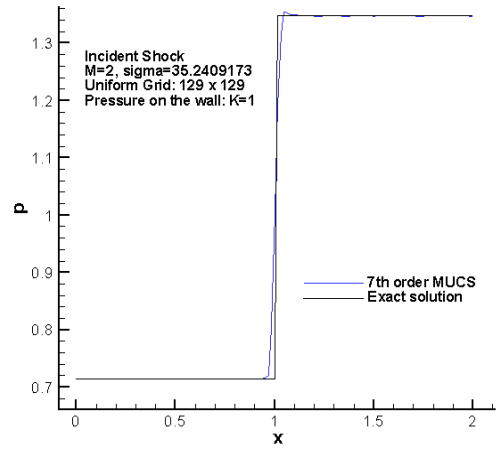

(b)

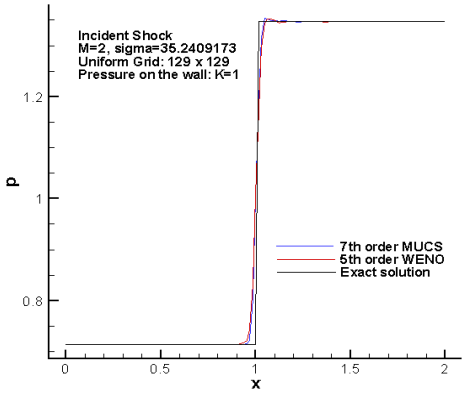

(c)

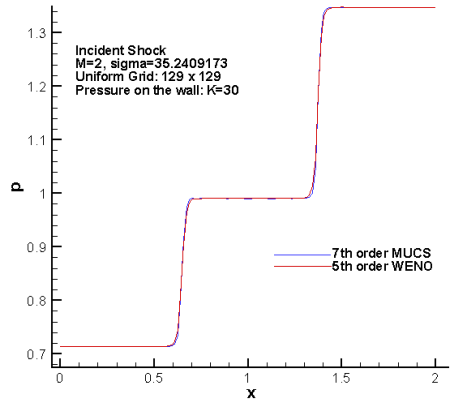

(d)

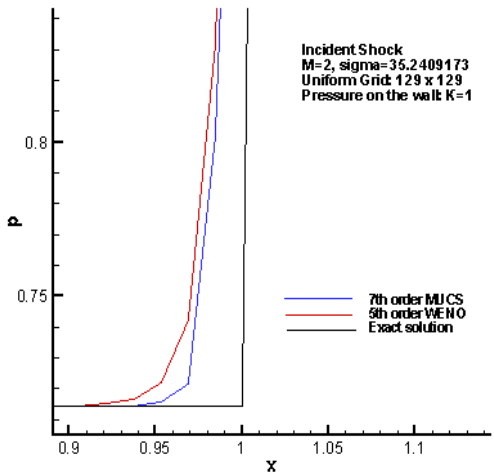

(e)

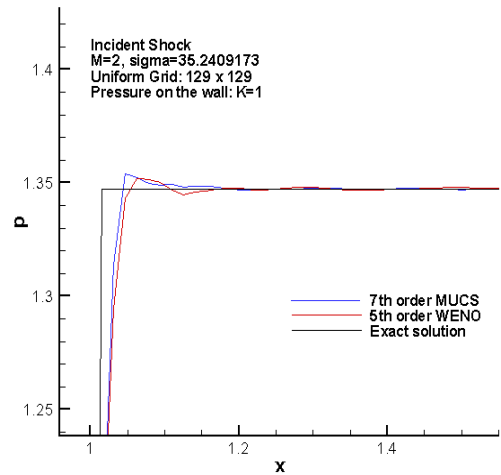

(f)

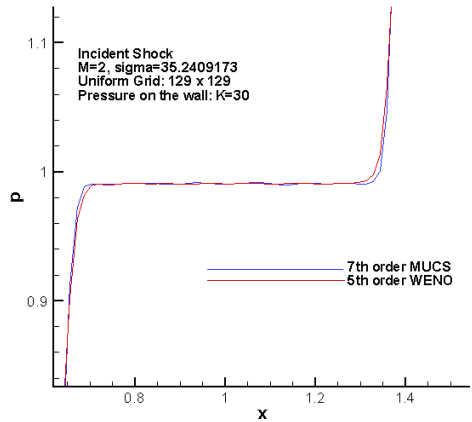

(g)

**Figure 4.3 Pressure distribution on fine grids (129x129)** (a) 7$^{th}$ order MUCS (b) MUCS and Exact
(c) 7$^{th}$ order MUCS , 5$^{th}$ order WENO and exact solution (d) Pressure on the level K=30
(e) (f) (g) Locally enlarged comparison

## V. Conclusion

1) MUCS with a new shock detector and new mixing function, which uses WENO to improve upwinding compact scheme, is ready to use for both 2-D and 3-D Euler and N-S for sharp shock capturing and high resolution for small length scales.
2) MUCS does not have case related parameters.
.

## Acknowledgments

This work is supported by AFRL VA Summer Faculty Research Program. The authors thank Drs. Poggie, Gaitonde , Visbal for their support through VA Summer Faculty Program.

## References

[1] Jiang, G. S., Shu, C. W., Efficient implementation of weighted ENO scheme. *J. Comput. Phys.,* 126, pp.202—228, 1996.
[2] Lele S.K., Compact finite difference schemes with spectral-like resolution, *Journal Computational Physics*, 103, pp.16—42, 1992.
[3] Liu, C. and Oliveira, M., Modified Upwinding Compact Scheme for Shock and Shock Boundary Layer Interaction, *AIAA Paper* 2010-723
[4] Ren, Y., Liu, M., Zhang, H., A characteristic-wise hybrid compact-WENO scheme for solving hyperbolic conservation laws, *Journal of Computational Physics* 192 (2003) 365–386, 2003